\documentclass[letterpaper]{article}
\usepackage{aaai19}
\usepackage{times}
\usepackage{helvet}
\usepackage{courier}
\usepackage{graphicx}
\usepackage{subfig}
\usepackage{multirow}
\graphicspath{ {./Figures} }
\begin{document}

%\title{A Machine Learning Approach for Quantitative Measurement of Relative Market Efficiency}
\title{Intra-day Equity Price Prediction using Deep Learning\\ as a Measure of Market Efficiency}
%\author{Anonymous Review Copy}
\author{David Byrd\thanks{David Byrd is a research scientist at the Georgia Institute of Technology in Atlanta, GA. (db@gatech.edu)},
Tucker Hybinette Balch\thanks{Tucker Hybinette Balch is managing director at J.P. Morgan AI Research, and professor on leave at the Georgia Institute of Technology. (tucker@cc.gatech.edu)}}
\maketitle

\begin{abstract}
In finance, the weak form of the Efficient Market Hypothesis asserts that historic stock price and volume data cannot inform predictions of future prices. In this paper we show that, to the contrary, future intra-day stock prices could be predicted effectively until 2009. We demonstrate this using two different profitable machine learning-based trading strategies. However, the effectiveness of both approaches diminish over time, and neither of them are profitable after 2009. We present our implementation and results in detail for the period 2003-2017 and propose a novel idea: the use of such flexible machine learning methods as an objective measure of relative market efficiency.  We conclude with a candidate explanation, comparing our returns over time with high-frequency trading volume, and suggest concrete steps for further investigation.

\begin{quote}
\end{quote}
\end{abstract}

\section{Introduction}

In March 2017, Ana Avramovic, an analyst at Credit Suisse, published an impactful whitepaper that was immediately reported on by the {\em Financial Times}, {\em Barrons}, {\em Marketwatch} and others  \cite{avramovic2017were}. She asserted that High Frequency Trading (HFT) has irrevocably changed the structure of equity trading markets in the United States. She described how even investors who may not realize it are involved in HFT: ``As institutional investors avail themselves of these sophisticated algorithms, and discount brokers fill retail trades through HFT wholesalers, we are all high frequency traders now.''

Avramovic shows that the volume of intra-day HFT grew substantially from 2003 to 2009. Her premise is that as the volume of HFT accelerated from 2000 to 2009, the markets became correspondingly more efficient. In order to support her premise, she shows how several measures of efficiency have changed over time. In particular, she offers historical bid/ask spreads (the difference between the highest bid for a stock and the lowest ask), and the distribution of volume in trading during the day as evidence of changing market efficiency. 

We revisit Avramovic's work, but with a new tool. We introduce a quantitative method for objectively measuring market efficiency: Namely by assessing how well basic Machine Learning algorithms can extract profit from price data in intra-day trading.  Higher profitability indicates reduced market efficiency, while reduced or zero profitability indicates improved market efficiency. 

Our study shows that significant profit could have been made intra-day using basic Machine Learning methods before 2009. This suggests that the market held exploitable inefficiencies at that time. After 2009 our reference algorithms are no longer able to profit from intra-day information, implying that the markets have become more efficient.

\section{Background and Related Work}
Concepts that eventually evolved into the Efficient Market Hypothesis can traced to Louis Bachelier, who first described price speculation as a``fair game'' which should have an expected return of zero \cite{bachelier1900theorie}.

In 1970 Eugene Fama laid out what he called ``an extreme null hypothesis'': The idea that ``security prices at any point in time fully reflect all available information'' \cite{malkiel1970efficient}. Over the years, tests of the hypothesis have settled into three general categories: Weak, semi-strong, and strong.  In this work we are focus on the weak form, for which the ``available information'' is simply a sequence of historical prices:
\begin{quote}
\textbf{Weak-form Efficient Market Hypothesis (EMH):} Future equity prices cannot be predicted by analyzing prices from the past.
\end{quote}

\begin{quote}
For the purposes of this study we consider the EMH to imply that if an algorithm {\it can} predict a future price and profit from it, the market is less efficient, while if that same algorithm cannot predict future prices the market is comparatively more efficient.
\end{quote}

In a later work Fama updated the classification of weak-form tests to include dividend yields and interest rates as information sources, and the target was generalized to ``return predictability''.  \cite{fama1991efficient}

Reliable links have been established between information efficiency of a market and its bid/ask spread.  In 2007, Tarun Chordia {\it et al} investigated the relationship between liquidity and market efficiency, concluding that ``liquidity stimulates arbitrage activity, which, in turn, enhances market efficiency'' and in 1986 Amihud and Mendelson wrote that ``Illiquidity can be measured by the cost of immediate execution'' and ``Thus, a natural measure of illiquidity is the spread between the bid and ask prices'' \cite{chordia2008liquidity,amihud1986asset}.  Together these results suggest negative correlation between market efficiency and bid/ask spread.

Market efficiency has a pragmatic implication for traders and portfolio managers. For them an efficient market means they can easily buy or sell a stock without incurring undue cost: An efficient market is indicated by small bid/ask spreads where the difference between the highest bid for a stock and the lowest ask price is minimized. Another indication of efficiency is consistent trading volume throughout the day versus a common situation where trading volume is exaggerated at the beginning and end of the trading day \cite{avramovic2017were}.

In a 2010 concept release, the United States Securities and Exchange Commission (SEC), characterized High Frequency Trading (HFT) activity as those practices that utilize sophisticated computation, co-location, short trading time windows, and frequent order cancellation. Such strategies produce a large number of trades each day.  The SEC noted that high frequency traders typically exit all of their positions (i.e., sell all of their stocks) at the end of the trading day to avoid potential risk associated with holding positions overnight \cite{sec2010concept}.

HFT activity can benefit traders in a number of ways, including:
\begin{itemize}
    \item To reduce the impact and cost of large individual transactions by breaking them into many hundreds or thousands of smaller transactions over a trading day. \cite{hendershott2011does}
    \item Profiting from very short term arbitrage opportunities in which a future stock price can be predicted from immediately available information. These approaches usually involve co-location in which the trader's computers are located at the exchange to reduce timing delays. \cite{angel2013fairness}  ``The speed of trading ultimately affects how profitable HFT strategies will be,'' is how Tim Klaus characterizes it. ``A thousandth or even a millionth of a second can make a difference.'' \cite{klaus2017market}
\end{itemize}

Exchanges (e.g., the New York Stock Exchange) incentivize some HFT behavior by offering rebates to traders who add liquidity by posting and filling limit orders, as explained by Thierry Focault: ``When a trade takes place, they [the exchange] charge a take fee to market takers and rebate part of this fee to market makers.'' \cite{foucault2013liquidity} The above cited SEC report found that HFT indeed adds liquidity to the market.  This conclusion is supported by a separate study by Albert Menkveld in which he shows that when a new HF Trader joined Chi-X Europe (a pan-European exchange) the benchmark bid/ask spread immediately decreased by 50\% in stocks accessible to the HF Trader vs those not accessible \cite{menkveld2013high}.

\begin{figure*}[t]
    \centering
    \subfloat[2003-2008]{
        \includegraphics[width=8cm]{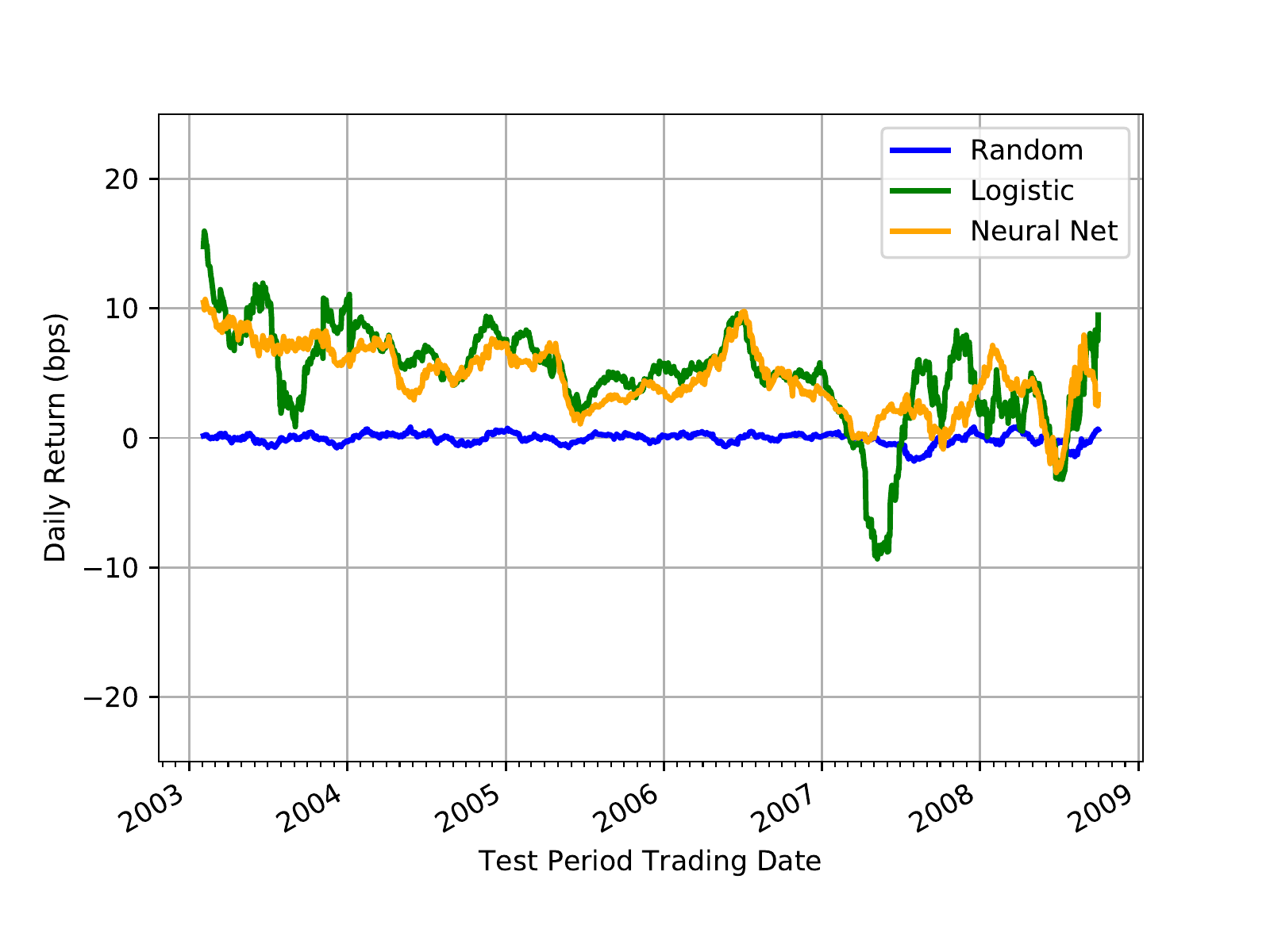}
    }
    \subfloat[2009-2017]{
        \includegraphics[width=8cm]{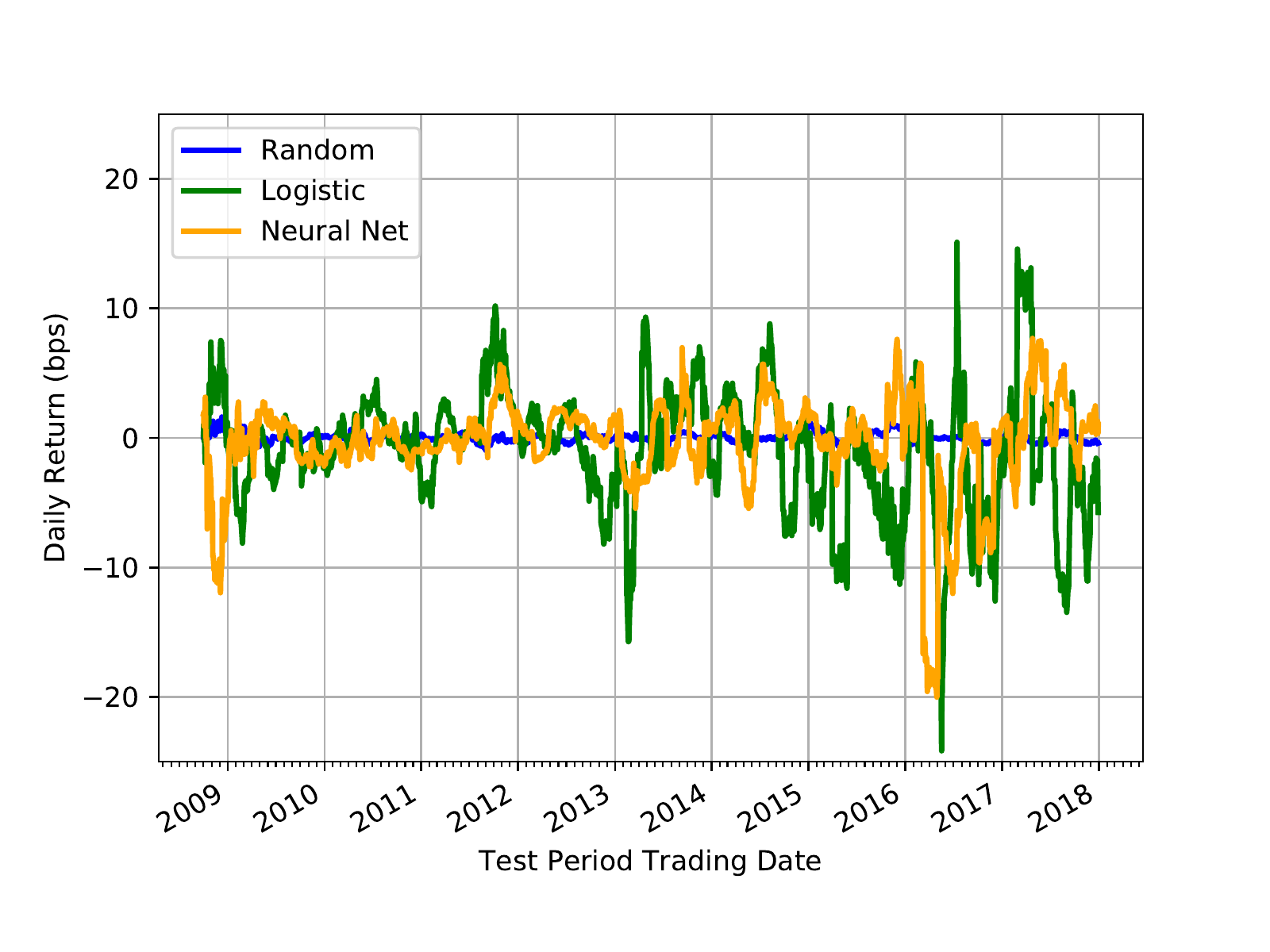}
    }
    \caption{Daily returns for three reference classifiers before and after peak daily volume.}
    \label{fig:split_daily}
\end{figure*}

\begin{figure*}[t]
    \centering
    \subfloat[2003-2008]{
        \includegraphics[width=8cm]{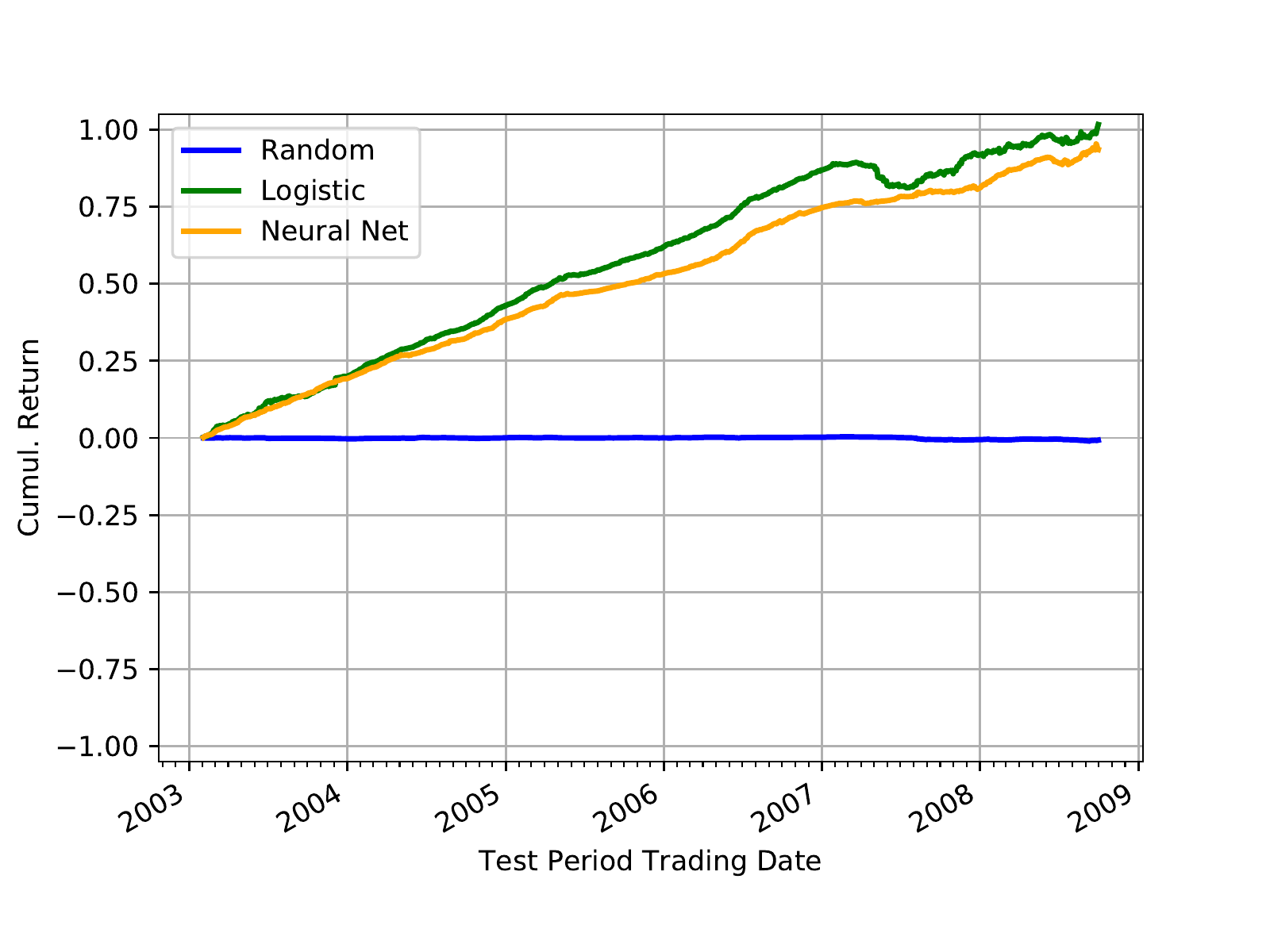}
    }
    \subfloat[2009-2017]{
        \includegraphics[width=8cm]{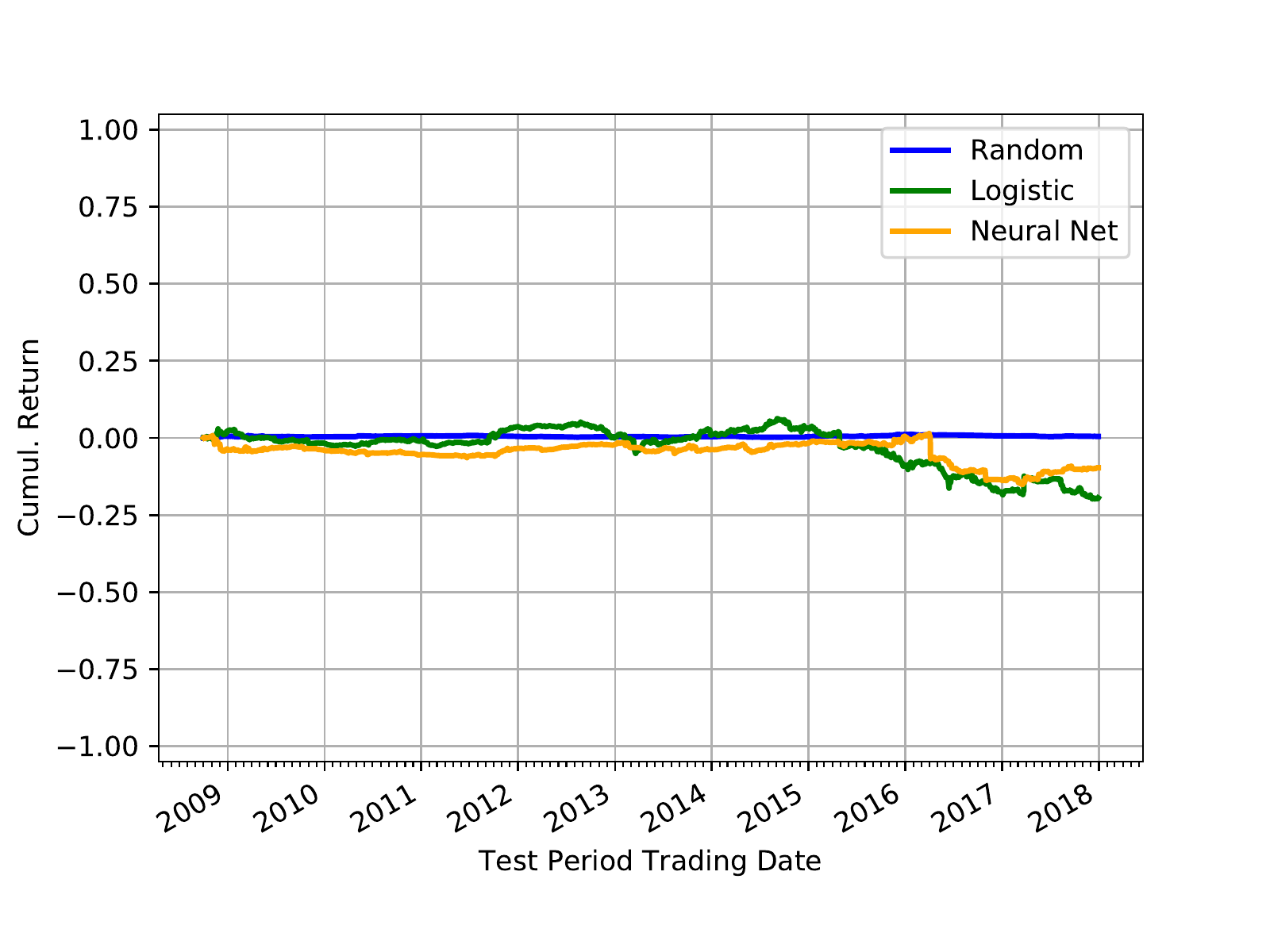}
    }
    \caption{Cumulative returns for three reference classifiers before and after peak daily volume.}
    \label{fig:split_cumul}
\end{figure*}

\section{Approach}

We created three reference trading systems, which we call {\it reference} systems because they are intended to be basic representatives of the corresponding learners -- not elaborate finely-tuned implementations.   Our aim is to compare relative performance of the same algorithms over time, not to ``beat the market''.

Each system includes two models: A {\it long} model that identifies situations when a stock is likely to go up in price  and should be purchased, and a {\it short} model that identifies situations where the stock price is likely to to down and the stock should be shorted (or bet against):
\begin{itemize}
    \item {\bf Neural Network (NN):} A basic neural net classifier that uses earlier intra-day price data to decide whether to buy a particular stock, short it, or do nothing.
    \item {\bf Logistic Regressor (LR):} A standard logistic regression model that makes classifications as described above for the NN.
    \item {\bf Random Classifier:} A randomized classifier to serve as a control.
\end{itemize}
All three systems are trained and tested with the same data sets and evaluated and compared on an equal footing.  We consider all stocks traded on all major US exchanges from 2003 to the end of 2017. The data is consolidated into one-minute units called ``minute bars'' for consideration by the learning trading systems.

{\bf Neural Network:} The first model employed in our trading strategy is a low-depth fully connected feedforward neural network with two hidden layers of size 180 and 20 using ReLu activation. The input size is determined by the \texttt{end\_x} parameter described in the hyperparameter section below, and the output layer contains two neurons with softmax activation.  The network is trained via adaptive momentum gradient descent (Adam) with a categorical cross-entropy loss function using early stopping against validation loss.  Each example observation is a series of one-minute close prices for a single equity up to the \texttt{end\_x} minute.  Each label is the market close price for the same equity transformed to a one-hot vector with True indicating the symbol moved in the target direction by more than some threshold hyperparameter.

{\bf Logistic Regressor:} The second model employed in our trading strategy is a binomial logistic regression with L2 regularization.

{\bf Random Classifier:} The third model employed in our approach is a simple random classifier that makes predictions arbitrarily but does follow the class balance of the training data. This model serves as a control.

In our experiments each learning method is used in a trading strategy as follows:
The  strategy invests an equal amount of funds to the long and short side each day.  Each symbol placed in the positive class of the \emph{up} classifier and the negative class of the \emph{down} classifier will be allocated a long position.  Each symbol placed in the positive class of the \emph{down} classifier and the negative class of the \emph{up} classifier will be allocated a short position.  Symbols placed in the negative class by both classifiers (``no opinion'') or the positive class by both classifiers (``conflict of opinion'') will not be allocated funds.  We require that the investments are always long/short balanced.  The same total dollar allocation will be made to the long and short sides of the portfolio, each side subdivided equally among the available symbols.  The balance requirement is absolute: If there are long predictions but no short predictions, the strategy will not trade at all.

\section{Data}

The source data for our implementation is the NYSE Trade and Quote (TAQ) data set from Wharton Research Data Services (WRDS) under academic license \cite{nysetaq}.  We use the daily trade files, which summarize reported trades from all major U.S. exchanges at one second resolution through 2014 and subsequently at one millisecond resolution.

The trade data is reported per time period, per symbol, per exchange.  We process the source data 
% using the Python library Pandas to produce a compressed dataframe per day containing the same trade information as the original, then further process those files 
to produce a data set of one-minute open-high-low-close (OHLC) bars plus volume for all symbols on each day.  A stock that does not trade during a given minute retains its earlier close price with zero volume.  We further maintain a list of known symbol changes for consistency. We exclude Exchange Traded Funds (ETFs) and test symbols from consideration,

For each day in a training, validation, or test period, one-minute OHLC bars are retrieved from disk for a selected universe of ``important" stocks.  By {\it universe} we mean the set of equities eligible for trading on a particular day.  The universe is determined on a daily basis as the most traded 500 stocks from the previous twelve months.  For example the selected universe for the first day of a training period beginning January 1, 2002 is the top 500 dollar volume equities of year 2001.  Only this daily selected universe of equities is used for training, validating, or testing the system, and only these symbols are considered by the trading strategy on a given day.

\begin{figure*}[t]
    \centering
    \subfloat[Daily Return: Neural Classifier]{
        \includegraphics[width=8cm]{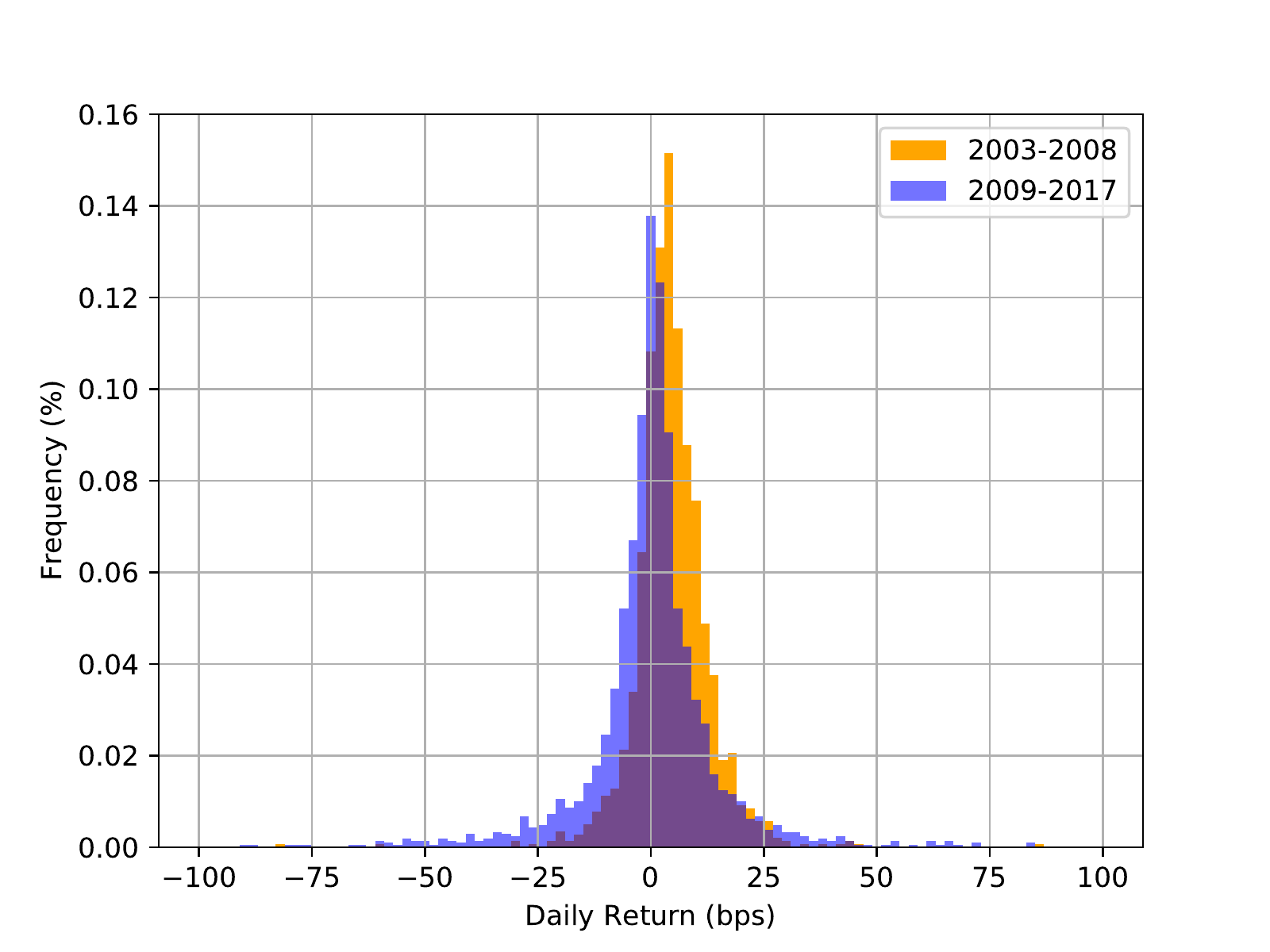}
    }
    \subfloat[Daily Return: Logistic Classifier]{
        \includegraphics[width=8cm]{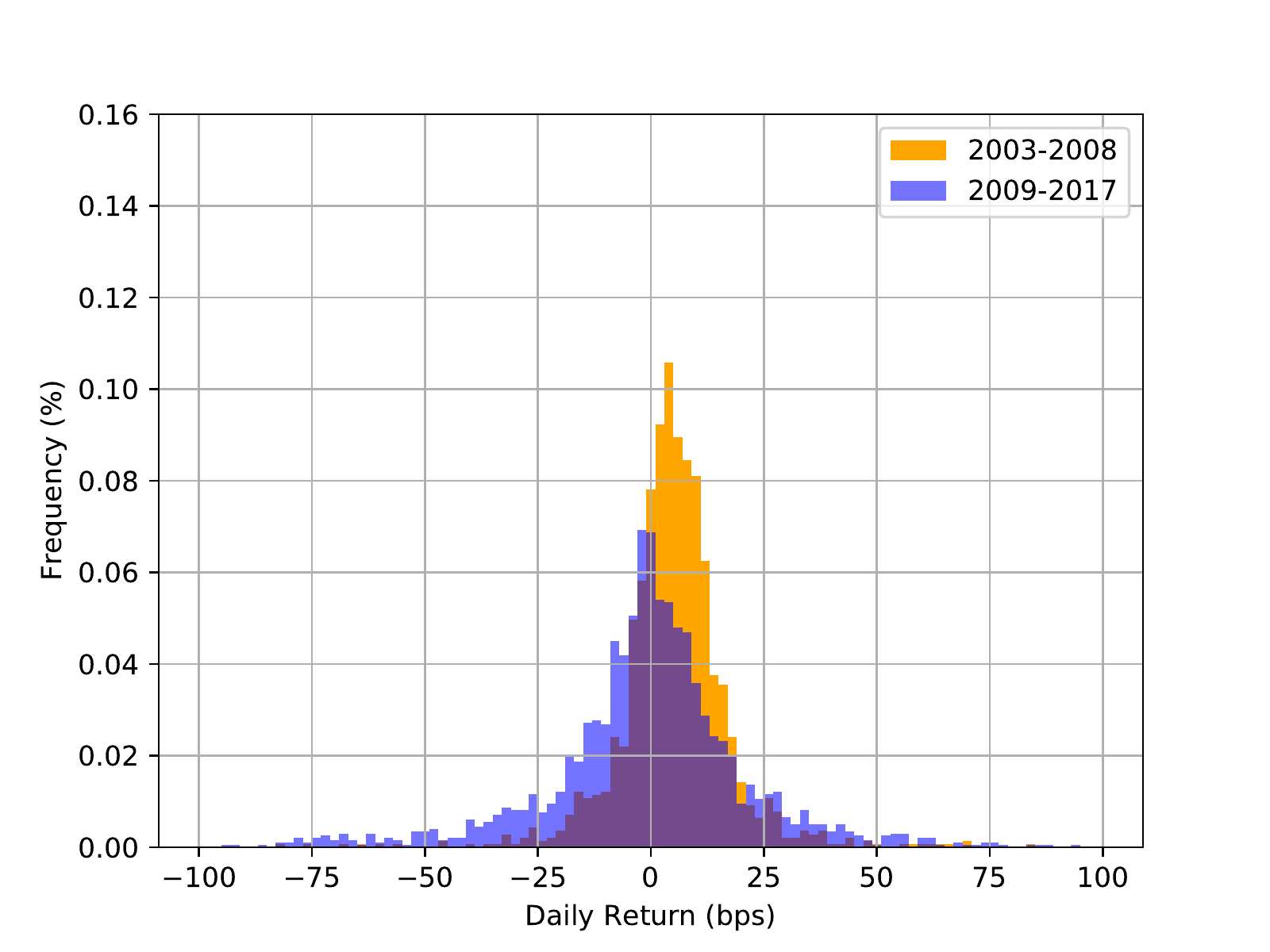}
    }
    \caption{Histogram of daily returns comparing same classifier across test periods.}
    \label{fig:histograms}
\end{figure*}

\begin{figure*}[t]
    \centering
    \subfloat[Trade Precision: Neural Classifier]{
        \includegraphics[width=8cm]{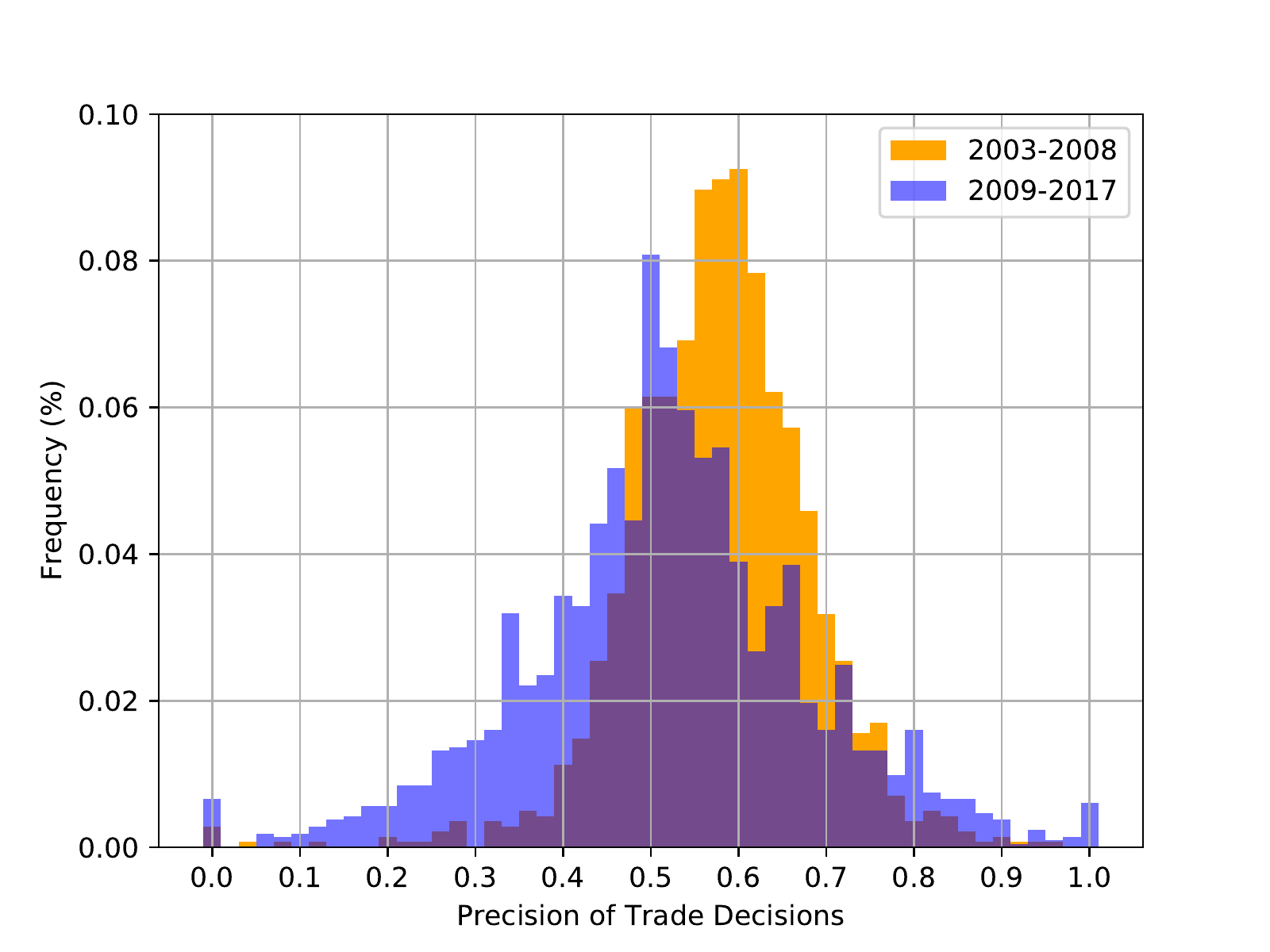}
    }
    \subfloat[Trade Precision: Logistic Classifier]{
        \includegraphics[width=8cm]{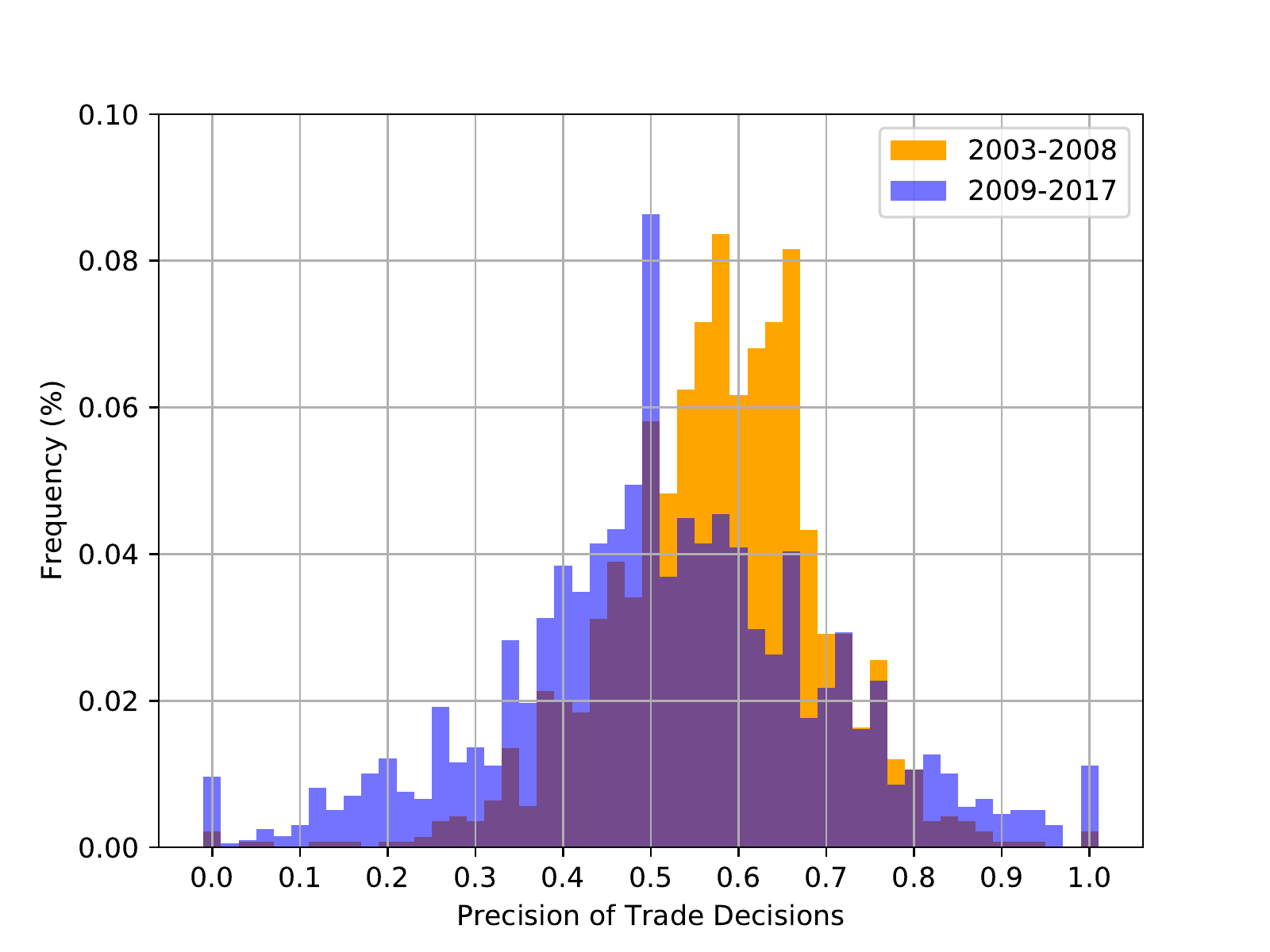}
    }
    \caption{Histogram of trade precision comparing same classifier across test periods.}
    \label{fig:prec_histograms}
\end{figure*}

\section{Methodology}

Given a time period over which we wish to conduct an experiment, we proceed in an iterative manner as follows.  The start date for our initial \emph{test} period is 25 months after the beginning of the experimental period.  For the current iteration, this date is treated as \emph{today} before market open, and no subsequent data may be used.  The \emph{validation} period is the month preceding the test period.  The \emph{training} period is the twelve months preceding the validation period.  The twelve months prior to the training period are needed for universe selection as previously described.

The system is trained and validated at the start of each month only.  For a single time iteration, the allowed universe of symbols is retrieved for every day during the training period.  These approximately $500 \times 252 = 126,000$ rows of 390 one-minute close prices become the available training data.  The same process is repeated for the validation and test periods.

{\bf Training Data:} Input observations are transformed into cumulative returns backward from a final observation minute \texttt{end\_x}.  Mean one-minute returns are also obtained for the universe and transformed in the same manner.  Finally, the difference between the individual observations and the selected universe is taken to produce universe-relative training data normed to 0.0 at minute \texttt{end\_x}.

For each input, the forward cumulative return is computed from minute \texttt{end\_x + 1} to market close, then made universe-relative as above to produce universe-relative gains from the end of the minute after the observations cease.  This permits one minute of time to make a decision and enter orders.  Norming two consecutive minutes to zero also produces a clean break that withholds any future information from the classifier.
  
\begin{figure*}[t]
    \centering
    \subfloat[Daily Return]{
        \includegraphics[width=8cm]{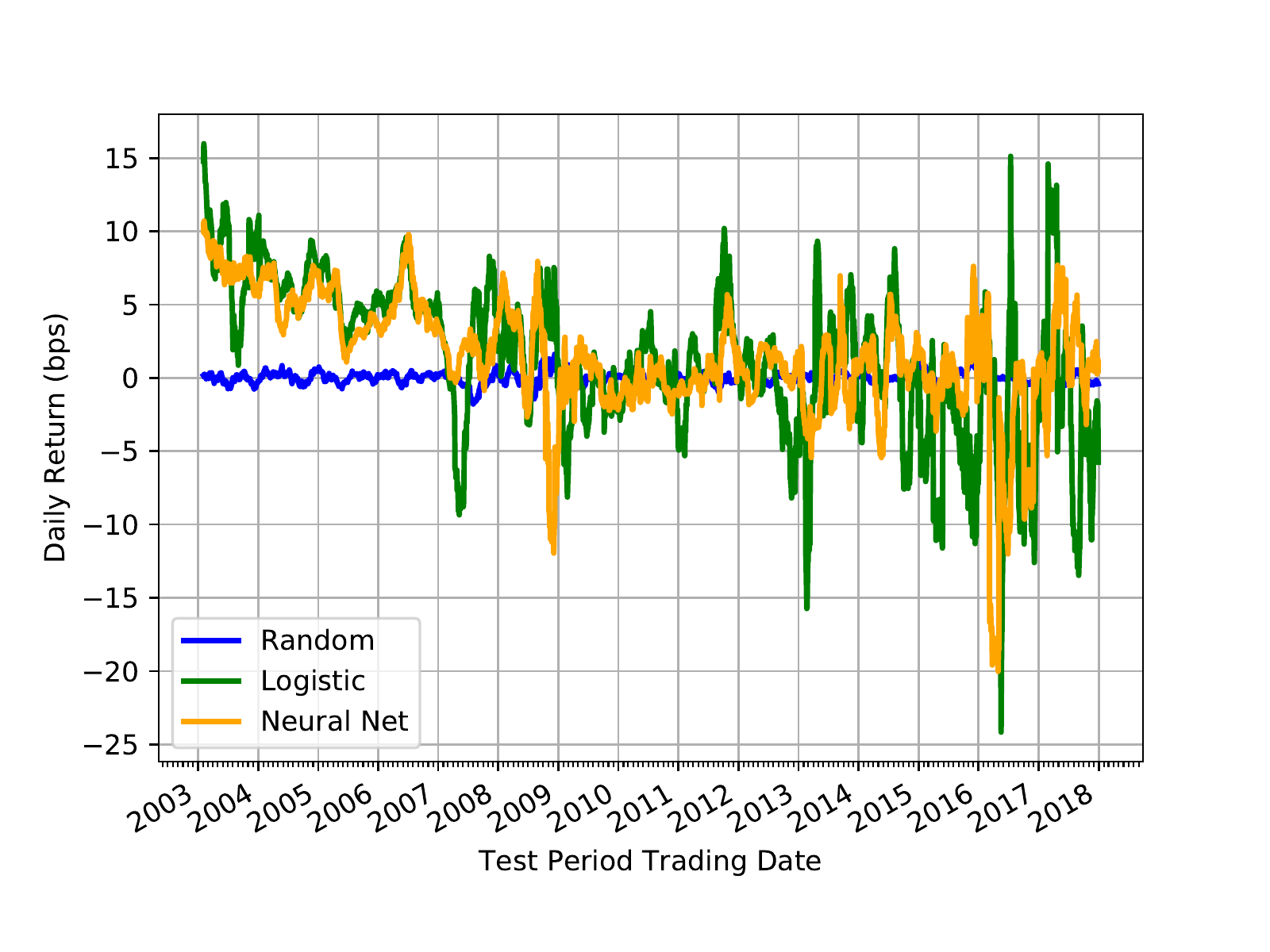}
    }
    \subfloat[Cumulative Return]{
        \includegraphics[width=8cm]{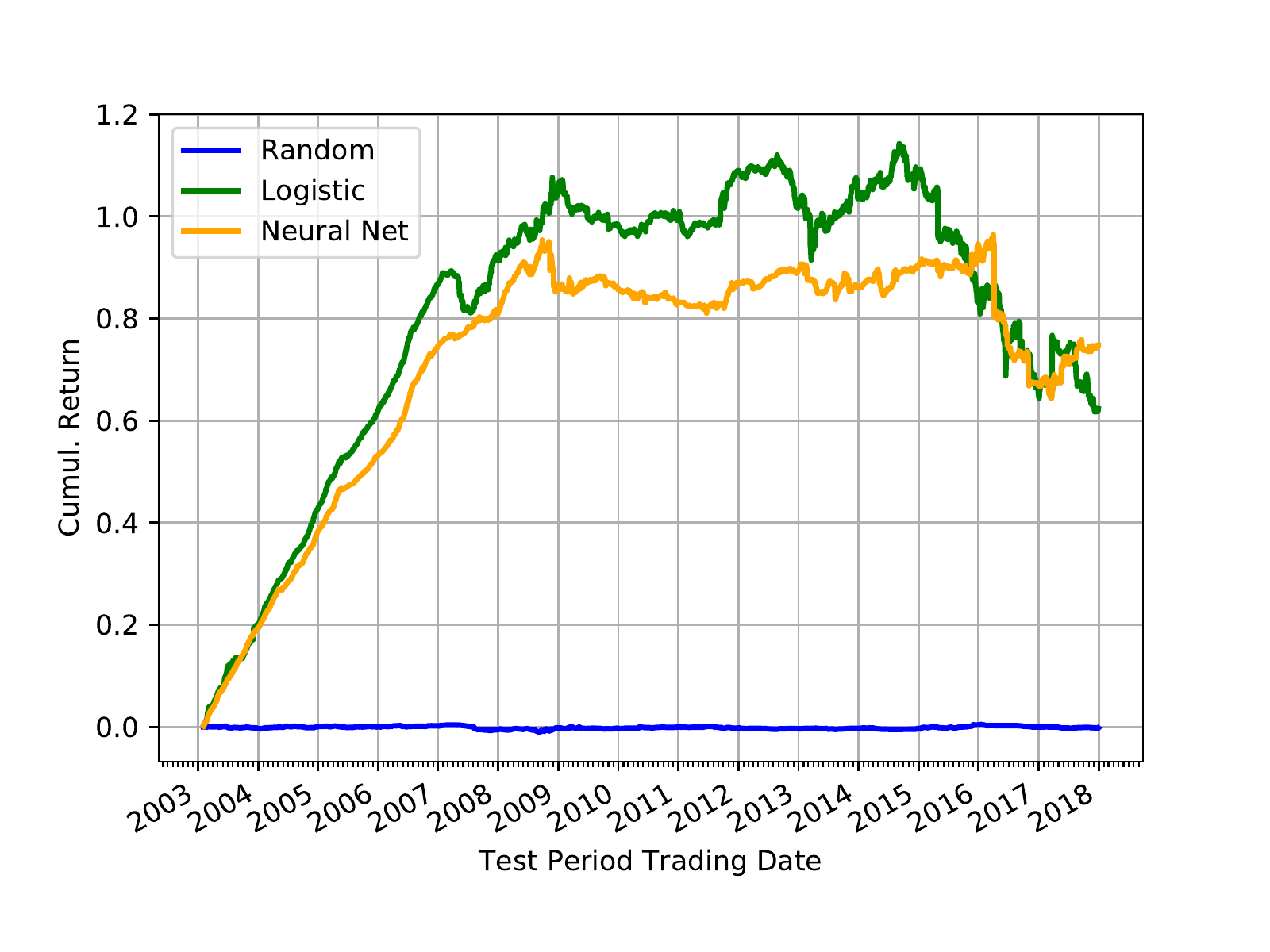}
    }
    \caption{Comparison of three classifiers over a fifteen year period.}
    \label{fig:returns_full}
\end{figure*}

\begin{figure*}[t]
    \centering
    \subfloat[Through Sep 2008]{
        \includegraphics[width=8cm]{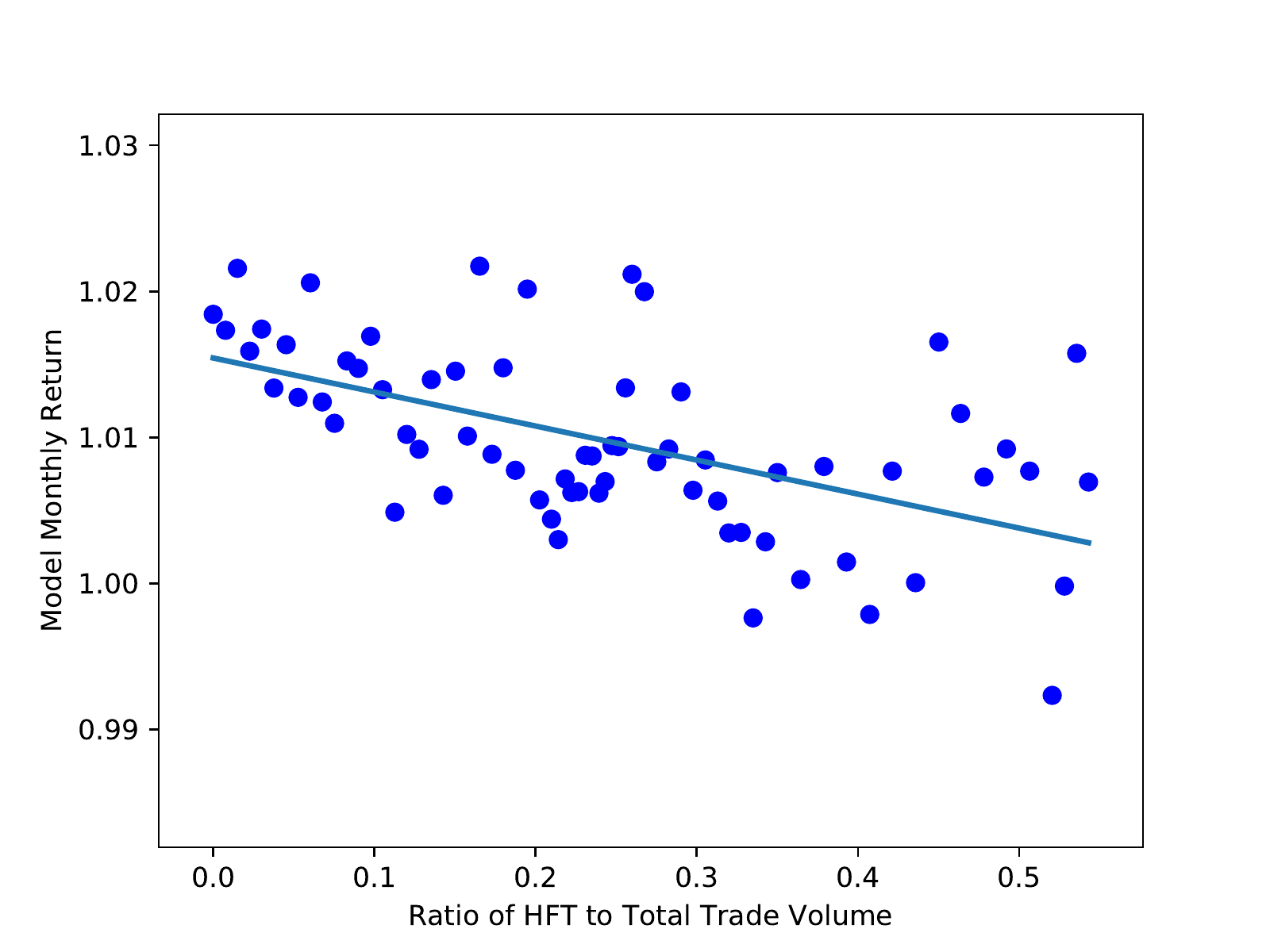}
    }
    \subfloat[Oct 2008 onward]{
        \includegraphics[width=8cm]{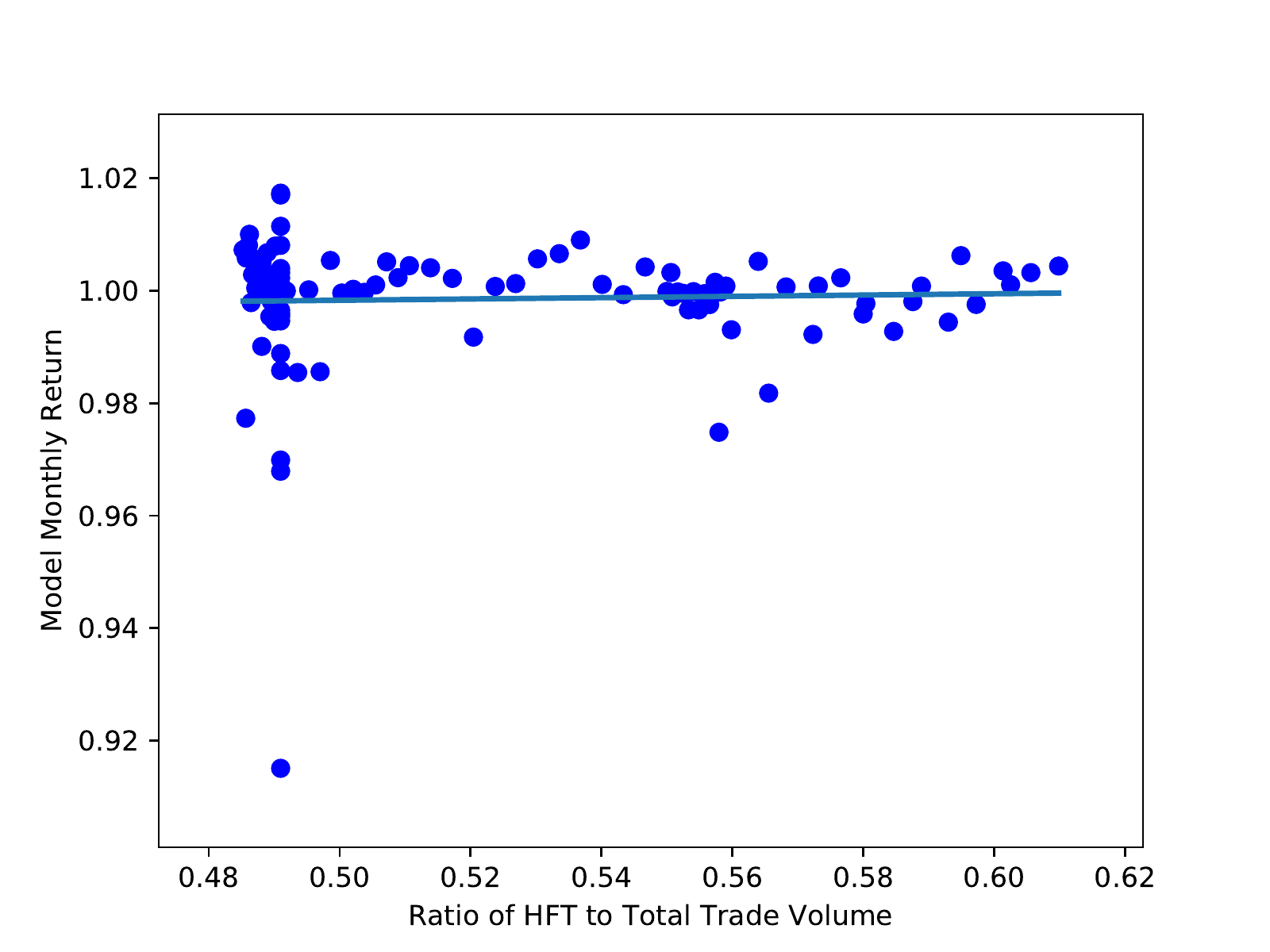}
    }
    \caption{Correlation between HFT Ratio and Model Return before/after volume peak.}
    \label{fig:hft_scatter}
\end{figure*}

{\bf Hyperparameter Search:} For each reference learner method two similar classifiers are trained to detect stocks that will rise or fall into the close.

The entry minute for the day's trades, and the threshold in bps (hundredths of a percent) required for a price movement to be considered meaningful, are hyperparameter choices for the system to make.  A hyperparameter grid search is performed across a simple $3 \times 4$ matrix: \texttt{end\_x} $\in \{ -5, -10, -30 \}$, \texttt{bps} $\in \{ 2, 5, 10, 25 \}$.  For each candidate hyperparameter pair, each classifier network is trained on all available training examples constructed as described above.  For each hyperparameter pair, the trained network is evaluated on the validation data.

The objective function for the hyperparameter grid search is the precision of the trading strategy, evaluated on a per-symbol basis, where the true positive class indicates the strategy allocated long funds to a symbol that did rise or allocated short funds to a symbol that did fall, and so forth.  The trained model with the maximum trading strategy precision for the validation month is selected for use during the upcoming test month.

{\bf Evaluation:} During the test month, each trained model is queried with the daily observations up to the selected \texttt{end\_x} and a trading decision is made for every symbol every day.  The results of the trading strategy are evaluated for precision, daily return, and cumulative return.

\begin{figure}[h]
    \centering
    \includegraphics[width=8cm]{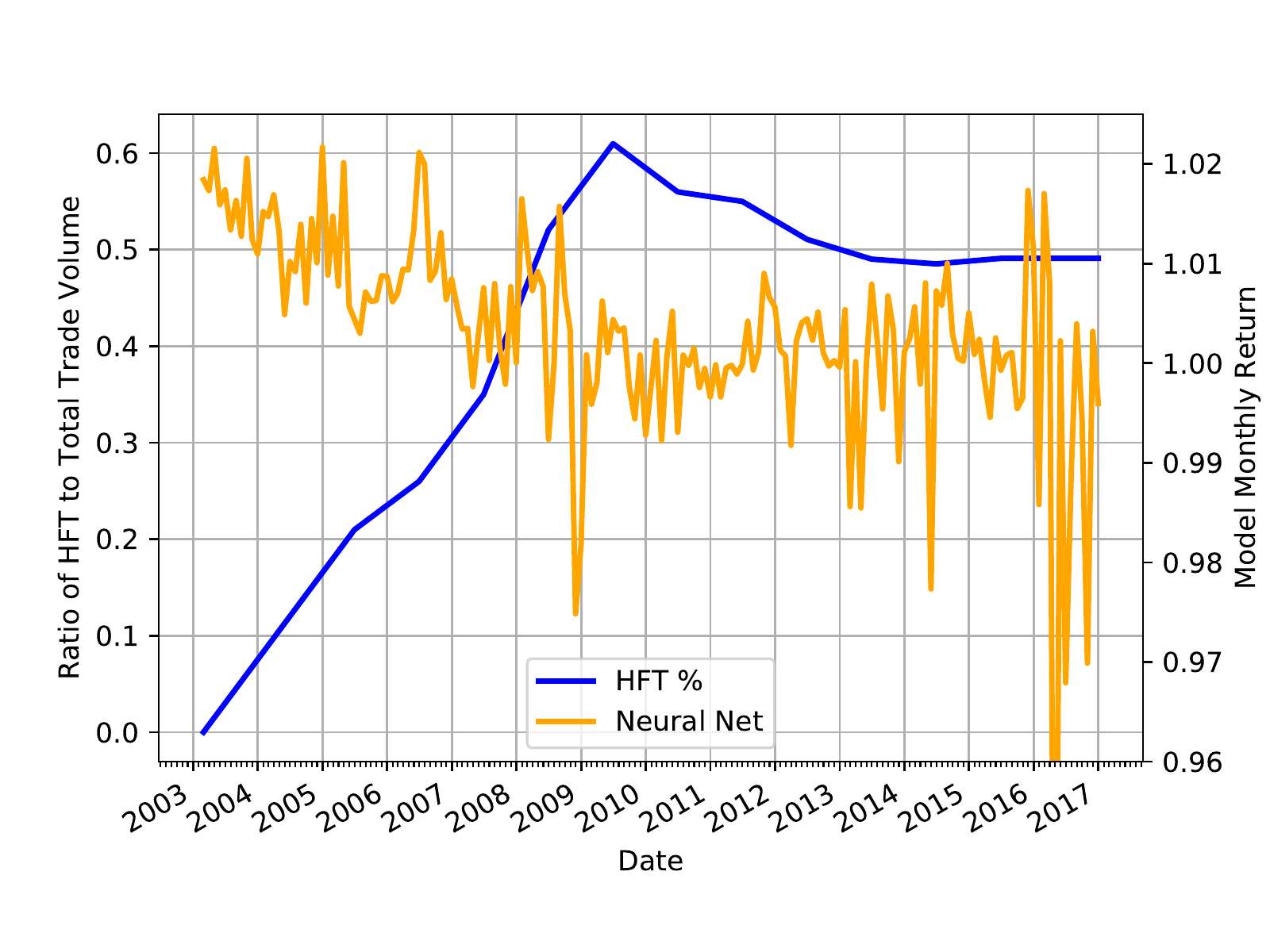}
    \caption{The percentage of High Frequency Trading has increased over time (blue line). Compare to the general decrease in the profitability of NN-based trading (orange).}
    \label{fig:hft_volume}
\end{figure}

The entire described training, validation, and testing process are repeated across the experimental period with the test period aligned to each calendar month in succession.  At the conclusion of the experiment, an analysis of logged models, trades, and results is used to produce and plot final daily and cumulative returns of the method across the entire period.

\section{Experimental Results}

In our initial analyses, we observed an abrupt change in market efficiency after October 10, 2008. In addition to that week representing one of the most significant drops in US market value, it was also the date of the highest ever intra-day volume of trading.  Accordingly, we selected the September/October 2008 boundary as a dividing point in our evaluation of market efficiency. We separately evaluate the data before October 2008 as ``early'' and after as `late.''

As mentioned above, we evaluated three types of classification models (neural network, logistic regression, and random classifier) over these two time periods: 2001-2008 and 2009-2017, with each model trained each month on a roll-forward basis.  Trade decisions are made on a daily basis using the most recently trained model.  

All models were trained from, and evaluated on, the same data using the same hyperparameter search method, on a monthly basis.  We compare results from the same classifier across time periods and across classifiers within the same time period.  All returns are quoted in basis points (bps) or one-hundredths of one percent.

{\bf Returns by Time Period:} The daily returns, cumulative returns, and precision of all three classification models are summarized in Table \ref{table:results}.  The returns of each model are illustrated in Figures \ref{fig:split_daily} and \ref{fig:split_cumul}.  As mentioned above, the analysis is broken into a ``before'' and``after'' period at September 30, 2008, near the point of greatest universe share trade volume (October 10, 2008).  Daily returns are smoothed with a 40-day centered window.

Observe that by all metrics the ML methods perform substantially better in the Early period, and furthermore that the control method (Random) is neutral in all time periods.

\begin{table}[h]
    \centering
    \begin{tabular}{ll|c|c|c|}
        & & Daily Ret & Cumul Ret & Precision \\
        \hline
        \multirow{2}{1.2cm}{Neural} & Early & 4.6 & 93.5 & 0.577 \\
                               & Late & -0.4 & -9.7 & 0.519 \\
                               \hline
        \multirow{2}{1.2cm}{Logistic} & Early & 4.9 & 101.6 & 0.576 \\
                               & Late & -0.9 & -19.5 & 0.519 \\
                               \hline
        \multirow{2}{1.2cm}{Random} & Early & 0.0 & 0.0 & 0.499 \\
                               & Late & 0.0 & 0.0 & 0.501 \\
                               \hline
    \end{tabular}
    \caption{Daily Returns (bps), Cumulative Returns (\%), and Precision for three reference classifiers for two time periods.}
    \label{table:results}
\end{table}

To further assess the changing effectiveness of each model over time, we considered the \emph{distribution} of daily returns for each model in the Early period and the Late period.  Figure \ref{fig:histograms} provides a histogram of returns for each learning algorithm using bins of two basis points.  In both cases, the distribution of returns in the early period is more narrow and centered in positive territory, while in the late period the distribution is broader and zero centered, suggesting that returns in the early period  are more positive and more reliable.  The logistic classifier, while providing slightly higher returns in the early period, has a wider variance of returns in both periods.

{\bf Accuracy of Trading Decisions:} We have primarily focused on the potential monetary returns of the reference classifiers across time, as is common in financial market applications.  In the field of Artificial Intelligence it is customary to evaluate classifier accuracy.  

Our method employs two classifiers together to make a single decision for each symbol in our universe each day.  Because no action is taken on the output of a single classifier, we do not report classifier precision.  Instead we report ``trade precision'', which is the precision of our system's decisions on each day the market is open.  

We present the distribution of daily ``trade precision'' in Figure \ref{fig:prec_histograms}.  Note that we do not assess false negative decisions (or ``recall'').  Our method allocates the same total dollar amount long and short regardless of the number of symbols selected, so there is little harm in omitting potentially-profitable symbols (Type II error), only in including unprofitable symbols (Type I error).  Because we evaluate the classifiers with universe-relative returns, we would expect a random classifier to have precision 0.5, and indeed it does.  Compared to this, both learning methods have substantially better than random discrimination and a relatively narrow distribution in the early time period, but not in the late time period.

Figure \ref{fig:returns_full} shows the daily and cumulative returns of all three model types across the entire test period from February 2003 to December 2017.  The initial performance of both learning algorithms is strong, though they appear to already be in decline as our test period begins.  (Earlier data was not available to us.)  Around the date of peak universe volume (October 10, 2008) both methods begin to produce noisy, flat returns.

\section{Discussion and Future Work}

We present objective measures of market efficiency, namely two reference machine learning trading methods that operate intra-day (Neural Networks and Logistic Regression).  We show that both of these trading strategies provide substantial profit from 2003 to 2008, but that their profitability degrades gradually over that time.  We further show that a Random Classifier control model trained and evaluated using the same process, method, and data has effectively zero returns throughout the entire study period.

This degradation in profitability for the same flexible learning model suggests a corresponding increase in weak-form market efficiency over the same period.  What it does not address is any underlying cause.

{\bf Relationship of HFT Volume to Weak Efficiency:} Recall that we assert that if a trading algorithm can profit using only historical price data, the market must be weak-form inefficient.  And conversely if that same algorithm is unable to find profit at a different time, we infer that the market is relatively more efficient. We concede that there likely exist other algorithms that can extract profit even when ours cannot.  Accordingly we only make claims regarding \emph{comparative} efficiency.

Consider the prevalence of HFT (see Figure \ref{fig:hft_volume}).   According to Avaramovic's data, there was almost no HFT volume in the US before 2003. The proportion of HFT out of all market volume accelerated from 2003 to 2009.  In the Introduction, we discuss Avramovic's claim that High Frequency Trading (HFT) has improved the efficiency of US markets.

To evaluate this claim, we compare the percentage of HFT volume in the market to the profitability of our trading algorithms. Figure \ref{fig:hft_scatter}  presents a scatterplot of the estimated monthly ratio of HFT volume \cite{avramovic2017were} to total US equity trading volume against the monthly returns captured by our neural learning model, separated at the time of peak universe volume.  Note that the time of our peak universe volume precedes the peak HFT volume by some months.  

To assess the data, we measured Pearson's correlation coefficient between the two series at -0.552 (strong negative correlation) for the early time period and 0.038 (no correlation) for the late time period.  Then in Figure \ref{fig:hft_volume} we overlay the same two series as a line plot.  The peak in HFT volume corresponds to the point at which our model returns flat-line. 

Thus it appears that this increase in efficiency corresponds to an increase in High Frequency Trading over the same period. Furthermore, after HFT volume peaks in 2008, our ML-based trading methods are ineffective. 

We acknowledge and address a shortcoming with this analysis: high-frequency trading has only been introduced to US markets once in history, with a corresponding single ``ramp up'' in volume.  While the correlation between efficiency (by way of reference ML model returns) and HFT volume ratio is strong, we could consider the entire rise of HFT to be a single event.  If the introduction of HFT to US markets for the first time could somehow be repeated, would this correlation hold over many such events?  Did HFT \emph{cause} improved efficiency?  These questions have no obvious answer using historical data.

To mitigate concerns over the $N=1$ nature of such a technological or methodological change in the market that cannot be repeated, we propose follow-on work: the careful implementation of an interactive simulated market in which numerous trading agents select and execute strategies from a flexible policy space, with analysis held until the market reaches competitive equilibrium, for example as used by the UMich Strategic Reasoning Group \cite{wellman2006methods}.  If we design an appropriate policy space, allow the market to reach equilibrium, and then add HFT strategies to the policy space, we can stochastically simulate as many independent introdictions of HFT as desired.  This more robust data set can then be subjected to the same experimental process we have here presented on historical data, to further validate or refute the conjecture that an increasing ratio of HFT volume improves market efficiency.

\section{Acknowledgements}

This material is based upon research supported in part by the National Science Foundation under Grant No. 1741026.

This material is based upon research supported in part by a J.P. Morgan AI Research Ph.D. Fellowship.

This paper has been prepared, in part, by the AI Research Group of JPMorgan Chase \& Co and its affiliates (“J.P. Morgan”) for information purposes, and is not a product of the Research Department of J.P. Morgan.  J.P. Morgan makes no explicit or implied representation and warranty and accepts no liability, for the completeness, accuracy or reliability of information, or the legal, compliance, financial, tax or accounting effects of matters contained herein.  This document is not intended as investment research or investment advice, or a recommendation, offer or solicitation for the purchase or sale of any security, financial instrument, financial product or service, or to be used in any way for evaluating the merits of participating in any transaction.   

\begin{quote}
\begin{small}
\bibliographystyle{aaai}
\bibliography{byrd_balch}

\begin{thebibliography}{}

\bibitem[\protect\citeauthoryear{Amihud and Mendelson}{1986}]{amihud1986asset}
Amihud, Y., and Mendelson, H.
\newblock 1986.
\newblock Asset pricing and the bid-ask spread.
\newblock {\em Journal of financial Economics} 17(2):223--249.

\bibitem[\protect\citeauthoryear{Angel and McCabe}{2013}]{angel2013fairness}
Angel, J.~J., and McCabe, D.
\newblock 2013.
\newblock Fairness in financial markets: The case of high frequency trading.
\newblock {\em Journal of Business Ethics} 112(4):585--595.

\bibitem[\protect\citeauthoryear{Avramovic}{2017}]{avramovic2017were}
Avramovic, A.
\newblock 2017.
\newblock We're all high frequency traders now.
\newblock {\em Credit Suisse Market Structure White Paper}.

\bibitem[\protect\citeauthoryear{Bachelier}{1900}]{bachelier1900theorie}
Bachelier, L.
\newblock 1900.
\newblock {\em Th{\'e}orie de la sp{\'e}culation}.
\newblock Gauthier-Villars.

\bibitem[\protect\citeauthoryear{Chordia, Roll, and
  Subrahmanyam}{2008}]{chordia2008liquidity}
Chordia, T.; Roll, R.; and Subrahmanyam, A.
\newblock 2008.
\newblock Liquidity and market efficiency.
\newblock {\em Journal of Financial Economics} 87(2):249--268.

\bibitem[\protect\citeauthoryear{Fama}{1991}]{fama1991efficient}
Fama, E.~F.
\newblock 1991.
\newblock Efficient capital markets: Ii.
\newblock {\em The journal of finance} 46(5):1575--1617.

\bibitem[\protect\citeauthoryear{Foucault, Kadan, and
  Kandel}{2013}]{foucault2013liquidity}
Foucault, T.; Kadan, O.; and Kandel, E.
\newblock 2013.
\newblock Liquidity cycles and make/take fees in electronic markets.
\newblock {\em The Journal of Finance} 68(1):299--341.

\bibitem[\protect\citeauthoryear{Hendershott, Jones, and
  Menkveld}{2011}]{hendershott2011does}
Hendershott, T.; Jones, C.~M.; and Menkveld, A.~J.
\newblock 2011.
\newblock Does algorithmic trading improve liquidity?
\newblock {\em The Journal of Finance} 66(1):1--33.

\bibitem[\protect\citeauthoryear{Klaus and Elzweig}{2017}]{klaus2017market}
Klaus, T., and Elzweig, B.
\newblock 2017.
\newblock The market impact of high-frequency trading systems and potential
  regulation.
\newblock {\em Law and Financial Markets Review} 11(1):13--19.

\bibitem[\protect\citeauthoryear{Malkiel and Fama}{1970}]{malkiel1970efficient}
Malkiel, B.~G., and Fama, E.~F.
\newblock 1970.
\newblock Efficient capital markets: A review of theory and empirical work.
\newblock {\em The journal of Finance} 25(2):383--417.

\bibitem[\protect\citeauthoryear{Menkveld}{2013}]{menkveld2013high}
Menkveld, A.~J.
\newblock 2013.
\newblock High frequency trading and the new market makers.
\newblock {\em Journal of Financial Markets} 16(4):712--740.

\bibitem[\protect\citeauthoryear{{New York Stock Exchange}}{2018}]{nysetaq}
{New York Stock Exchange}.
\newblock 2018.
\newblock {New York Stock Exchange Trade and Quote (TAQ) Data}.
\newblock Obtained via Wharton Data Research Services at
  https://wrds-web.wharton.upenn.edu/wrds/.

\bibitem[\protect\citeauthoryear{{United States Securities and Exchange
  Commission}}{2010}]{sec2010concept}
{United States Securities and Exchange Commission}.
\newblock 2010.
\newblock {Concept Release on Equity Market Structure, SEC Rel. No. 34-61358}.
\newblock Accessed via https://www.sec.gov/rules/concept/2010/34-61358.pdf.

\bibitem[\protect\citeauthoryear{Wellman}{2006}]{wellman2006methods}
Wellman, M.~P.
\newblock 2006.
\newblock Methods for empirical game-theoretic analysis.
\newblock In {\em AAAI},  1552--1556.

\end{thebibliography}
\end{small}
\end{quote}

\end{document}